\documentclass[letterpaper, 10 pt, conference]{ieeeconf}  % Comment this line out if you need a4paper

\IEEEoverridecommandlockouts                              % This command is only needed if 
\usepackage[pdftex]{graphicx}
\usepackage{caption}
\usepackage{subcaption}
\usepackage{url}
\usepackage{float}

\graphicspath{{./}}
\DeclareGraphicsExtensions{.png,.jpg,.eps,.pdf}

\author{Chuhao Liu and Shaojie Shen%
\thanks{All authors are with the Department of Electronic and Computer Engineering, Hong Kong University of Science and Technology, Hong Kong, China. {\tt\footnotesize $\{$cliuci, eeshaojie$\}$@ust.hk}}%
}

\title{\LARGE \bf
An Augmented Reality Interaction Interface for Autonomous Drone
}

\begin{document}
%To do
%- explain head gaze better
%- explain wireless communication better
%- Figure 1(a): change a map
%- extend to new users performance after training
%- some definitions: AR/MR interface, 3D map, projector, virtual drone
%- present user study better

\maketitle
\thispagestyle{empty}
\pagestyle{empty}

%%%%%%%%%%%%%%%%%%%%%%%%%%%%%%%%%%%%%%%%%%%%%%%%%%%%%%%%%%%%%%%%%%%%%%%%%%%%%%%%
\begin{abstract}

Human drone interaction in autonomous navigation incorporates spatial interaction tasks, including reconstructed 3D map from the drone and human desired target position. Augmented Reality (AR) devices can be powerful interactive tools for handling these spatial interactions. In this work, we build an AR interface that displays the reconstructed 3D map from the drone on physical surfaces in front of the operator. Spatial target positions can be further set on the 3D map by intuitive head gaze and hand gesture. The AR interface is deployed to interact with an autonomous drone to explore an unknown environment. A user study is further conducted to evaluate the overall interaction performance.

\end{abstract}

%%%%%%%%%%%%%%%%%%%%%%%%%%%%%%%%%%%%%%%%%%%%%%%%%%%%%%%%%%%%%%%%%%%%%%%%%%%%%%%%
\section{INTRODUCTION}

In recent years, drones are frequently used in industrial applications, such as space exploration after disasters, searches for missing people in the forest and facilities inspection. In these tasks, the environments are so complex that human manual control is risky or impossible. So, operators tend to use drones in an autonomous navigation mode, that the drone can perceive the environment and plan the trajectory by itself. The desired method is that a human operator interacts with the drone remotely by sending high-level commands, such as target flying position. However, there are various spatial interactions in autonomous navigation, including displaying the generated 3D map and the drone's position, as well as recognizing human-desired 3D tasks. These interactions are hard to complete on traditional flat-screen interfaces, such as computers or tablets. However, reality devices provide an immersive experience for spatial interaction. They are seen as the best option to interact with an autonomous navigating drone.

In this work, we develop a new AR interface to interact with our existing autonomous drone. Our drone can reconstruct a 3D map showing its surrounding environment, and this map is further displayed on the AR interface. On the 3D map, the user can set a flying target point by hand gesture with head gaze, which is normally hard to accomplish on a 2D screen interface. Thanks to the spatial mapping features in the AR device, the user can render the 3D map on the floor or table in front of him/her. The entire system provides an immersive and intuitive interaction with a drone during autonomous navigation. Meanwhile, a websocket-based broadcast program is applied to transmit data between the drone and AR interface. So that the interface can continuously interact with the autonomous navigating drone.

\begin{figure}[ht]
    \centering
    \begin{subfigure}{0.35\textwidth}
        \centering
        \includegraphics[width=\textwidth]{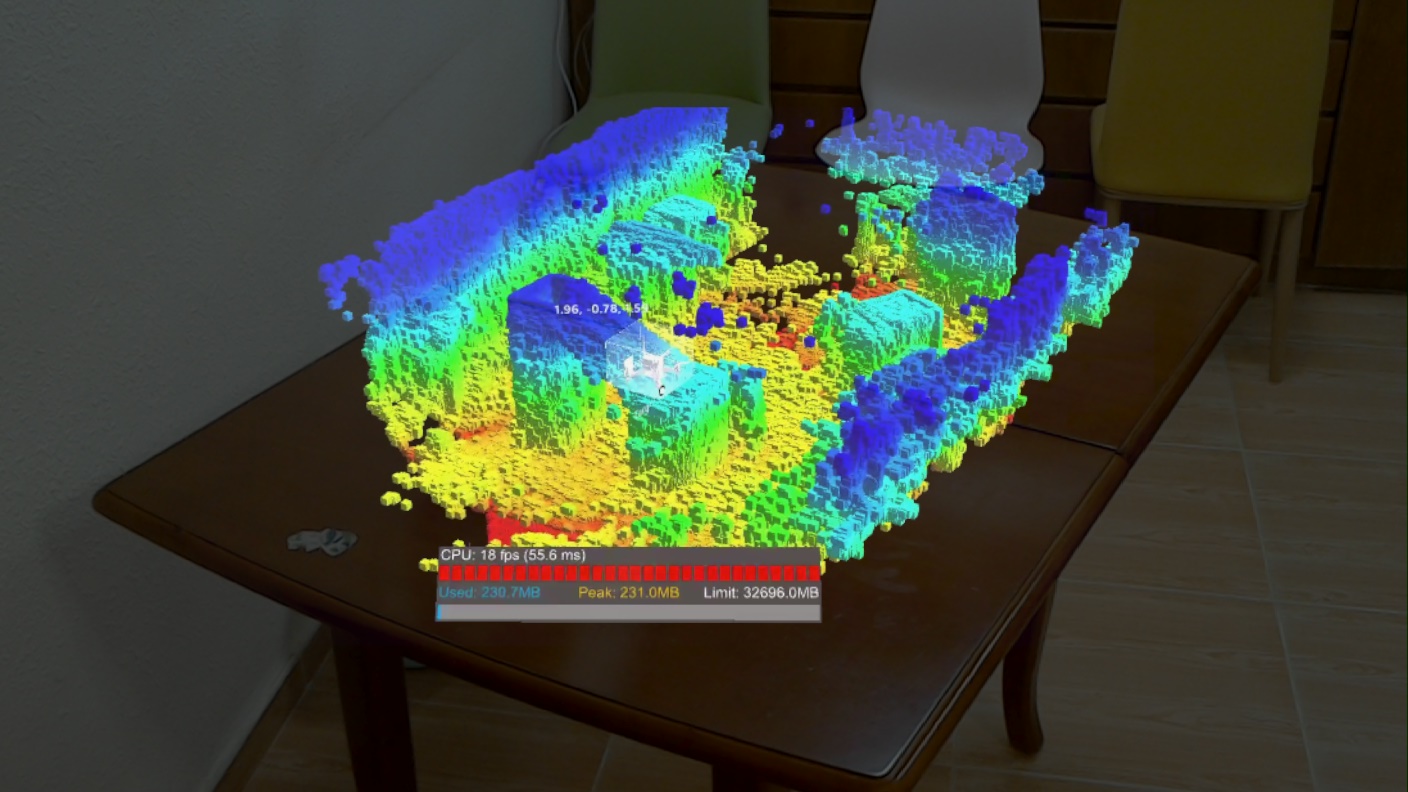}
        \caption{}
    \end{subfigure}
    \vfill
    \begin{subfigure}{0.35\textwidth}
        \centering
        \includegraphics[width=\textwidth]{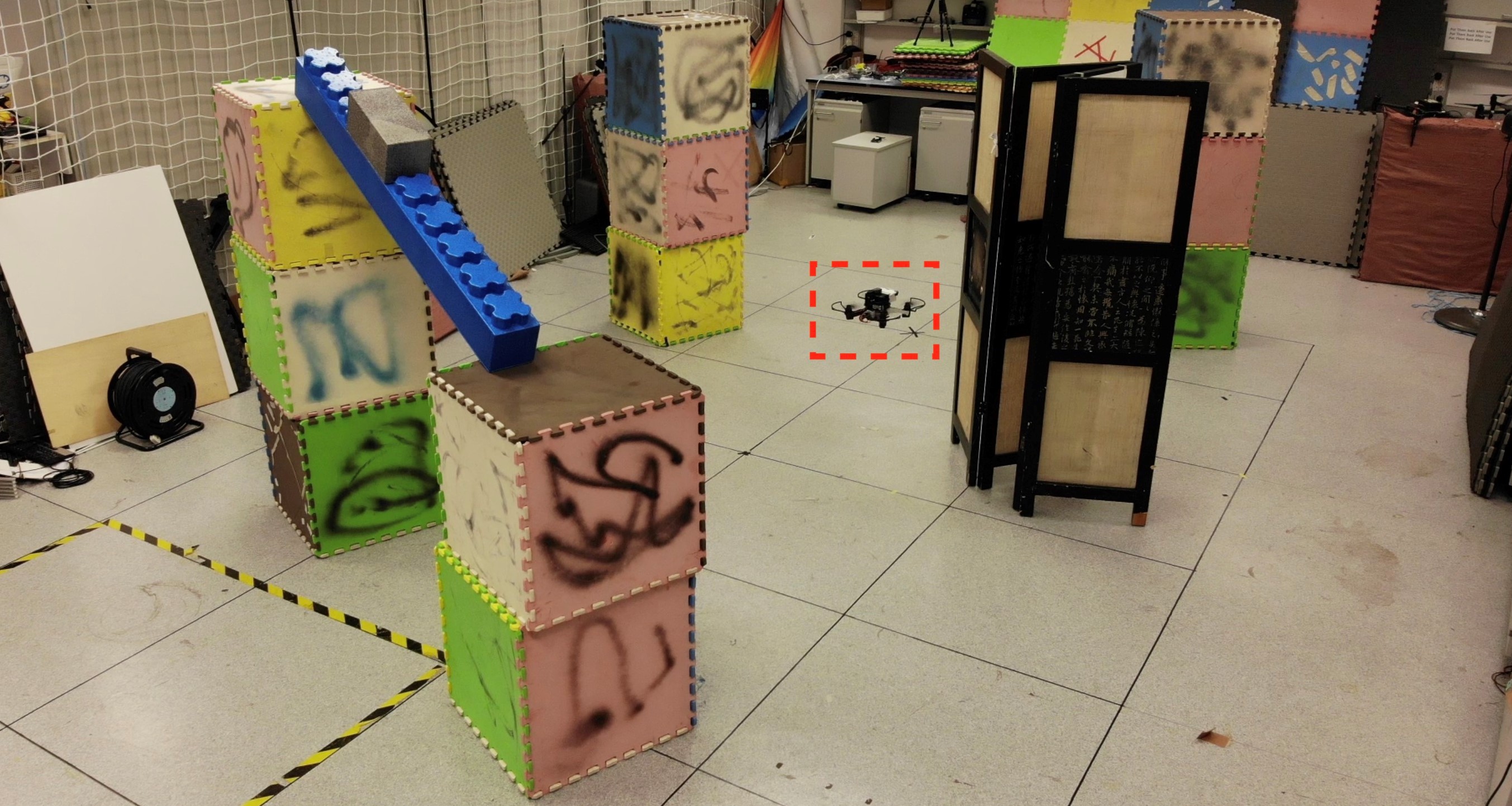}
        \caption{}
    \end{subfigure}
	\caption{(a) The AR interface during interaction with an autonomous flying drone. The 3D occupancy map is rendered on a physical table. The target position is represented by the white drone. A 3D target position can be set by intuitive manipulating the white drone. (b) The physical drone is flying autonomously in an indoor flight scene.}
	\label{mrintro}
\end{figure}

This article introduces our AR interface, which is based on Microsoft HoloLens 1. HoloLens is a head-mounted Mixed Reality (MR) device. However, we only use its AR functions, which render virtual objects without occlusion. Thus, we define our interface as an AR interaction interface. 
%However, we believe this is a general interaction approach and can be applied to either AR and MR devices as well.
The entire system is used in an indoor space exploration mission and further evaluated by a systematic user study. We conclude our main contributions as follows:

\begin{enumerate}
\item The drone's perceived 3D environment is rendered in scale on the real-world surfaces near the operator. The drone's real-time 3D position is also shown on the map. All the virtual objects are displayed immersive.
\item On the displayed environment, spatial targets are set by hand gesture and head gaze.
\item The AR interaction interface is deployed in an actual autonomous exploration mission and its overall performance is evaluated. A separate user study is conducted to evaluate interaction performance.
\end{enumerate}

\section{Related Works}

% \subsection{Traditional Human-Drone Interaction}
The traditional human-drone interaction (HDI) approach uses flat-screen interface and joystick controller to interact with drones. Some works have used a touch screen device to set target points easily \cite{kang2018flycam} \cite{lan2017xpose}. For example, an innovative interface in \cite{lan2017xpose} is deployed on an iPad to interact with an autonomous flying drone. It allows users to select desired photo-taking objects on the first-person-view image and the drone can collect the object image autonomously. For visualizing 3D spatial data, Rviz \cite{kam2015rviz} is a powerful computer graphic user interface (GUI). However, the traditional interface suffers from limitations in interacting with spatial objects.

% \subsection{VR, AR and MR interfaces. Interaction within visual range}
In spatial interaction tasks, reality devices have obvious advantages in rendering and conveying human intention. Earlier work in this area uses VR interface to display 3D point cloud map \cite{berge2016generation}, which is reconstructed offline by an airplane lidar. It shows the advantage of reality devices in visualizing a 3D map. To help human operator better understand the flying environment and setting target position, AR interface has also been used to display virtual drone and targets \cite{walker2019robot} \cite{walker2018communicating}. It guides the operator to joystick control a flying drone. The interface achieves good result in conveying drone's motion intent while the drone is in direct visual range of the operator. Another MR interface allows interaction with a flying drone by 3D rendering and voice control \cite{huang2019flight}. It renders a previous grid map on the flight space where has ground tag for drone's localization, and command the target grid by voice. The grid map it uses has only 4x4 resolution and request manual alignment with tag images on the ground. These works focused on using reality device to assist joystick-controlled fly. They are limited in scenarios that drone and operator shares a physical scene.

On the other hands, some works focus on using reality devices to interact with mobile robot remotely. A MR interface projects the real swarm drones as virtual drones \cite{honig2015mixed}, and operator further controls the virtual drones to reduce risk in face-to-face interaction with drones. Since this work does not have reconstructed 3D mapping from drones or human operators, the interface cannot help the operator understands the spatial map of the drone. Another good work \cite{stotko2019teleoperation} uses VR interface to assist remote tele-operation of a ground vehicle. The ground vehicle is joystick-controlled to explore an unknown environment and reconstruct the 3D space. Rather than displaying the first-person-view (FPV) image from the robot, the reconstructed 3D map is rendered on the VR interface to allow operator understand the spatial environment remotely. 

Among the reality devices used in above works, all of them have achieved immersive rendering. But differences exist between them that VR displays a pure virtual world, AR directly overlays graphics on the physical world \cite{milgram1994taxonomy}, and MR renders virtual objects in the physical world with occlusion. Based on performance in previous works, we choose to build our interface on AR device. Because VR totally blocks operator's visual information in the physical environment, and occlusion seldom happens in our scenario, AR becomes the best option for our system.

% Both of these two works make good start in using MR interaction on the flying drone. But they have limitations that require human operators to enter the flight space to gain direct spatial information. And the drones they use have limited autonomous flying function. 
%Innovative work in \cite{yuan2019human} used the human eye's gaze to direct a drone's flying, which achieves good results in real flight. However, eye gaze can only decide 2D position. It does not come out with a promising solution in deciding depth information.

% they focusses on assisting manual flying drone or drone with limited autonomous level. There are no odometry or rebuilt map from the drone to interact with.  

Unlike previous works, our interface focuses on interaction with autonomous navigating drone, which is beyond visual range of the operator. Real-time visual odometry and reconstructed 3D map are involved in the interface. To the best of our knowledge, this is the first Augmented Reality interface that interacts with an autonomous navigation drone in real-world flight. 

\section{Autonomous Drone System}\label{section_drone}

\begin{figure}[ht]
    \centering
    \includegraphics[width=\linewidth]{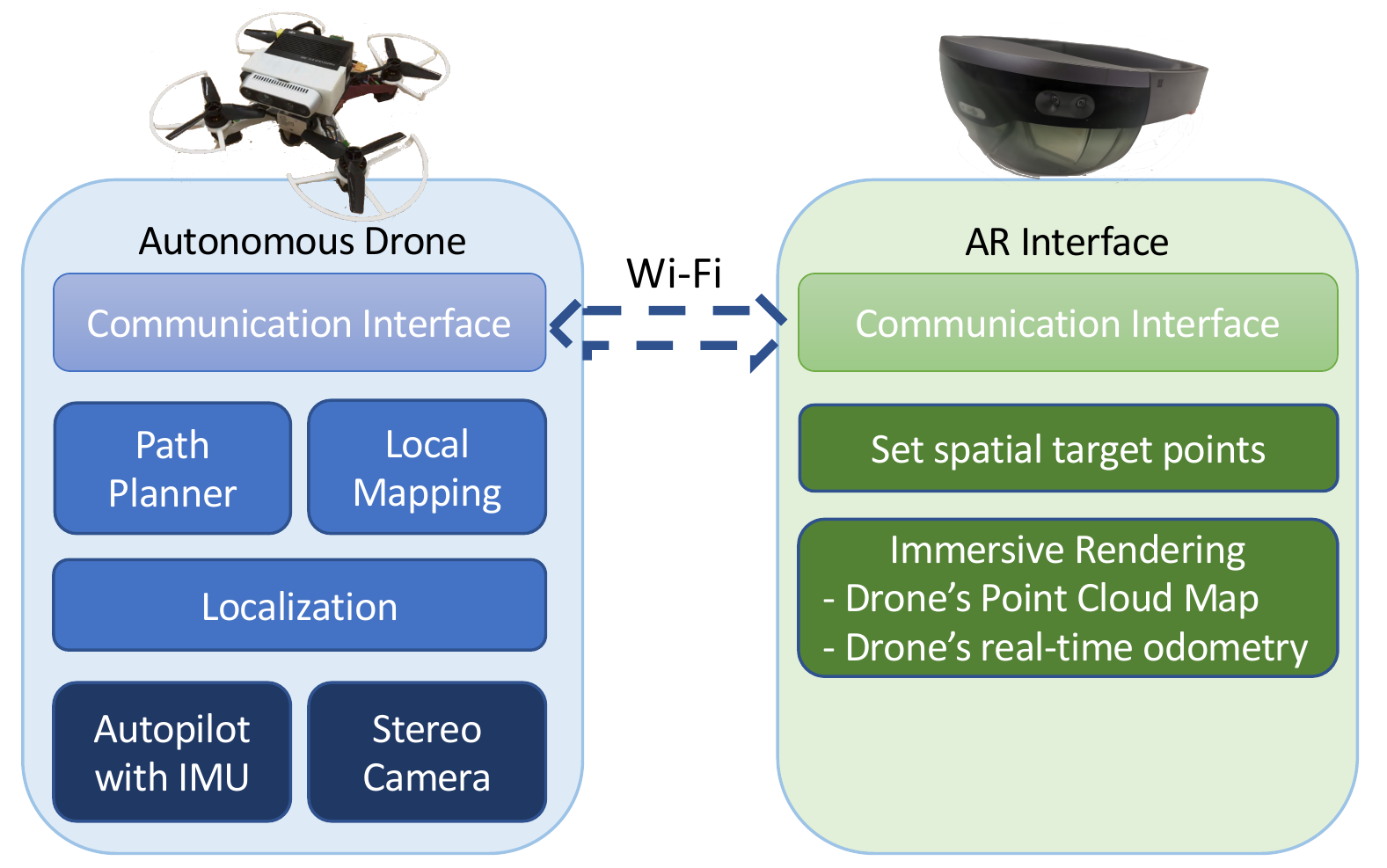}
    \caption{ Overall structure of the system: The autonomous drone has localization, mapping and path planner module running on the onboard computer. The AR interface can render a immersive 3D map and set spatial target positions. Occupancy map, VIO and flight commands are transmitted through Wi-Fi.}
    \label{system}
\end{figure}

The drone is equipped with a stereo camera, flight controller and onboard computer to run navigation-related modules.
VINS-Fusion \cite{qin2017vins}, a robust and accurate multi-sensor state estimator, is used for self-localization. It provide a robust and stable visual-inertial odometry (VIO). Pose graphs are optimized upon loop closure and can be saved as files for future use.

Based on the VIO from VINS-Fusion and depth images, Fast Incremental Euclidean Distance Fields (FIESTA) \cite{han2019fiesta} maintains a local occupancy map by raycasting. The occupancy map use cubic volumetric-elements (voxels) to represent the 3D structure around the drone. It is used for human visualization as well. FIESTA further generates an Euclidean signed distance field (ESDF) map, which calculates the Euclidean distance to the closet obstacle of each voxel and can be used for path planning. With ESDF map and VIO, Fast Planner \cite{zhou2019robust}
% a robust and efficient quadrotor motion planning system for fast flight in 3D complex environments, 
is applied to generate a safe, kinodynamic and smooth trajectory. It is able to read the target position and generate trajectory autonomously. The autonomously navigating drone is tested in various dense flying spaces.

After the mission finishes, Surfel Fusion \cite{wang2019surfel} is used offline to generate a detailed 3D mesh map from the recorded camera images and VIO. Surfel Fusion maintains a series of surfel maps with attached pose graphs to ensure global consistency.

% \subsection{Communication Interface}
The ROS platform is used to deploy all the navigation algorithms onboard. A communication module, rosbridge\footnote{http://wiki.ros.org/rosbridge-suite}, is further used to transmit data between the drone and AR interface by web-socket. 
% The wireless communicate is built on Wifi and communication protocol is further offered 
% using rosbridge \footnote{http://wiki.ros.org/rosbridge-suite}. 

\section{Interaction Approach}
We use Microsoft HoloLens 1 as the AR device in this work. The interface is developed on Unity 2018.4 and Visual Studio 2019 platform. HoloLens has its own visual SLAM algorithm running onboard \cite{hololens}. It provides a very robust and accurate VIO. The surrounded space is reconstructed and surface planes are recognized, such as table, ground and wall. No external sensors are required for the SLAM system of HoloLens. Besides its SLAM system, we choose HoloLens 1 because of its head gaze detection, gesture understanding, and convenient software development platform. 

\subsection{Rendering occupancy map}
\begin{figure}[ht]
    \centering
    {\includegraphics[width = 0.7\columnwidth]{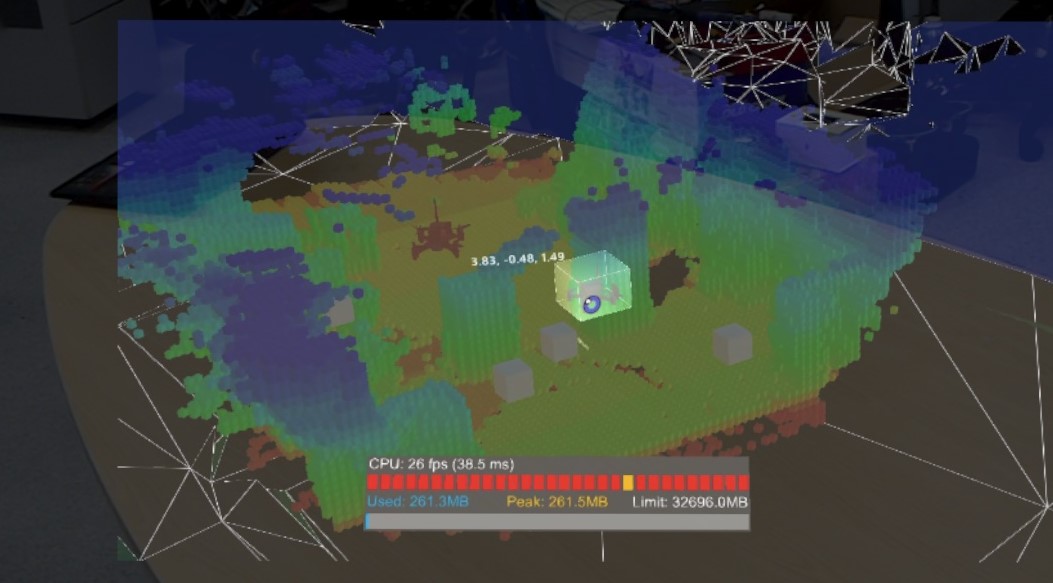}}
	\caption{\label{pcdmap} The occupancy map is rendered on a table in real-time. The red drone is drone's real position, which is updated continuously. The white drone is highlighted by a bounding box while pointed by head gaze. Then, operator can drag it by hand gesture to set a target position in the 3D spatial environment. Several grey cubes are placed on the map to indicate pre-designated targets for the test in user study.}
	\label{fig:renderpcd}
\end{figure}

In this system, the drone can navigate in an unknown environment and generate an occupancy map in real-time. On the map, the drone's recognized obstacles are represented by cubic voxels. The drone update the occupancy map at 10 Hz. To save computing resources on HoloLens, our AR interface updates the map message at 2 Hz, with a maximum of 30,000 voxels contained in each message. The interface continuously receives occupancy map and renders it through the GPU instancing function \cite{unitygpuinstance}, which allows HoloLens to render multiple cubic voxels in one function. It reduces the frequency of draw calls during the actual experiment and thus saves the computing resources of HoloLens. 

To initiate the rendering, the user can finger tap a physical point $O_{w}$ to set it as the origin of the rendered objects. On the display, the 3D map is scaled down to give the operator a clear navigation view. The $i^{th}$ voxel  has rendering position in the physical world,
\begin{equation}{\label{pcdtrans}}
    P^{w}_i = (sR_{d2w}P^{d}_i )+  O_{w}
\end{equation}
where $P^{d}_i$ is the position of the $i^{th}$ voxel in the occupancy map, $s$ is a map scaling factor, $R_{d2w}$ is the rotation matrix which can used to adjust the yaw angle of the rendered map, and $O_w$ is the physical origin of the rendered objects. During the interaction, the operator can change the scale factor $s$ and yaw angle of the rotation matrix $R_{d2w}$ to fit his or her own intuitive sensing. 
%Special notice that rotating and scaling the map cannot affect the rendering performance on physical surfaces.

Since the occupancy map does not have color information, we extract the height of each voxel to decide its RGB color. Thus, color of the 3D map gradually changes from the ground to the top. On this local map, we further display a virtual quadrotor to represent the drone's real-time odometry, as shown in Fig. \ref{pcdmap}. Then, the operator can monitor the drone's real pose without physically entering the flight scene. 

The AR interface ensures the rendered representatives are steadily placed in the real world. On a normal computer, although the 3D scene can be displayed, the point of view needs to be manually changed to recognize the 3D structure from a 2D screen. However, the rendered scene on our AR interface allows the operator to directly understand spatial objects.  The operator is also free to move his or her body to change a point of view intuitively.

\subsection{Render with spatial mapping}

\begin{figure}[ht]
    \centering
    \begin{subfigure}[b]{0.23\textwidth}
        \centering
        \includegraphics[width=\textwidth]{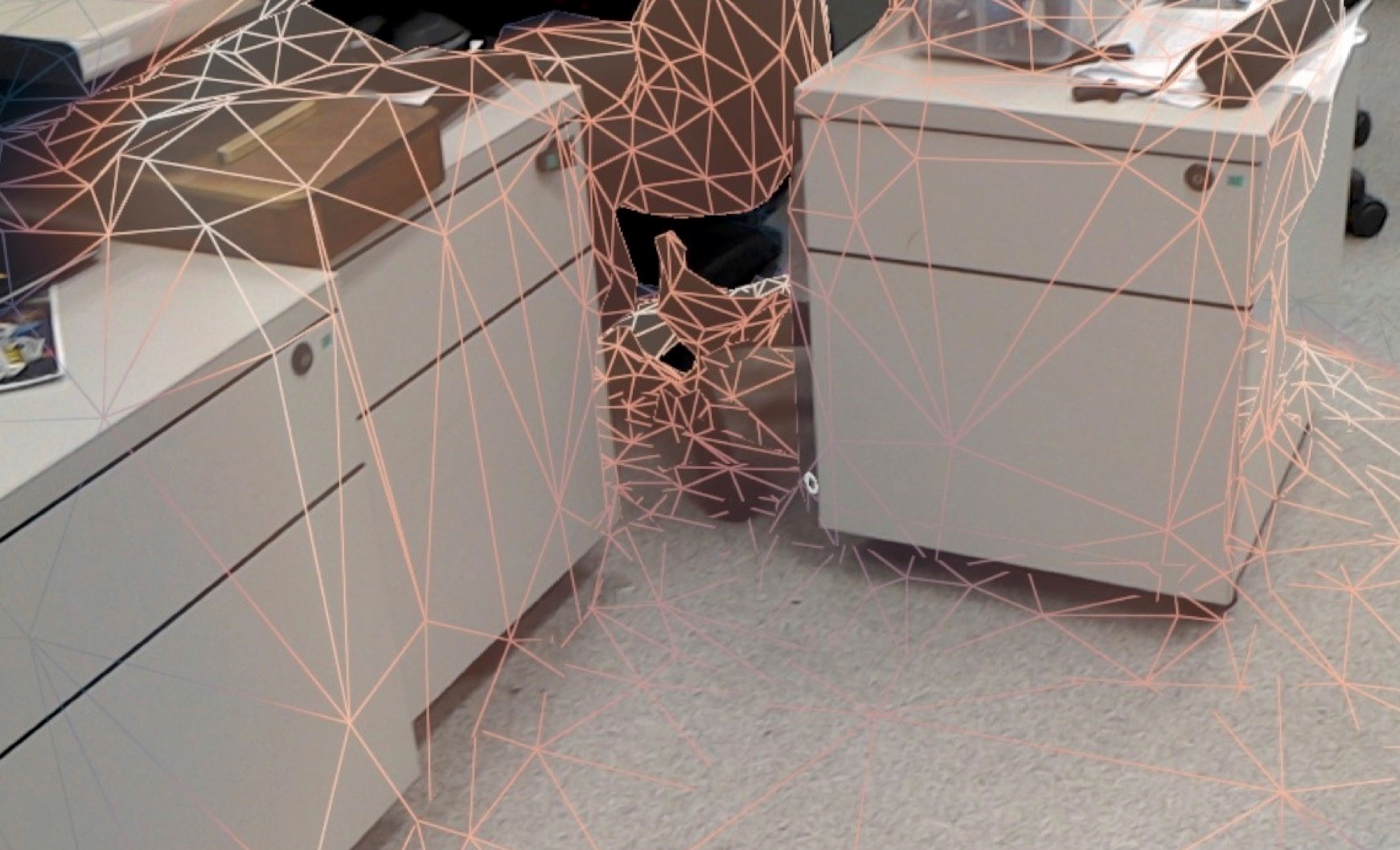}
        \caption{}
    \end{subfigure}
    \hfill
    \begin{subfigure}[b]{0.23\textwidth}
        \centering
        \includegraphics[width=\textwidth]{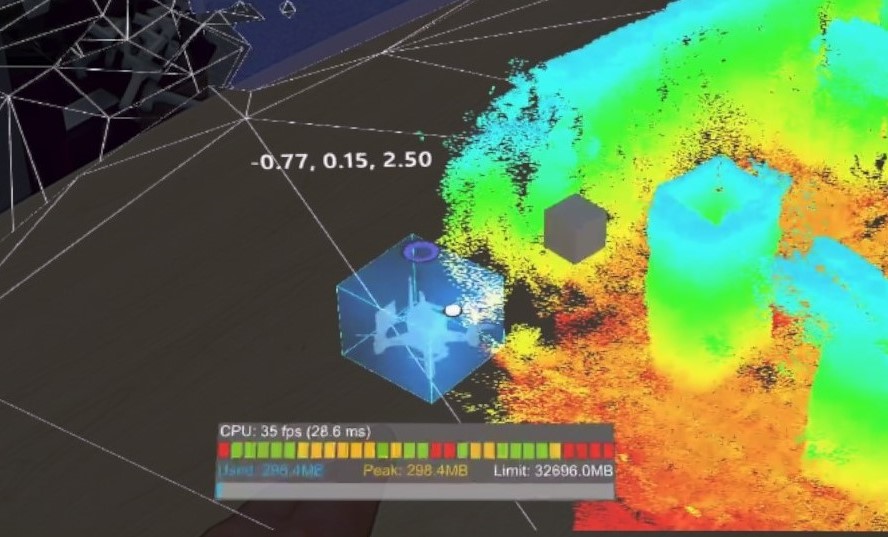}
        \caption{}
    \end{subfigure}
    
	\caption{(a) Spatial mapping: HoloLens scans one corner of a laboratory space. It extracts meshes (in white triangles) to represent objects and further converts surface meshes to planes. (b) Spatial manipulation: A head-directed cursor is pointed at the white drone. After the virtual drone is selected by finger tap, it is highlighted by a blue bounding box.}
	\label{fig:spatialmap}
\end{figure}

By taking advantage of AR devices, we can render the map at the recognized surfaces around the operator. So, the operator enjoys an immersive view on drone's rebuilt environment without losing information to his/her physical world. This function is based on spatial mapping of HoloLens. It first scans the environment and finds the surrounding meshes, as the white triangles shown in Fig. \ref{fig:spatialmap}(a). Then, the program generates planes based on the scanned mesh. Triangle meshes that fall within plane boundaries are removed in this step. The result shows the ground, walls and tables are recognized.

Once the surrounding surfaces are recognized, the operator can select a physical point $O_w$ as the origin of the rendered objects. $O_w$ is decided by head gaze and hand gesture. In our AR interface, a cursor overlays physical objects and continuously follows the head gaze of the operator. To select $O_w$, the operator moves the cursor to a desired point and selects it with a finger tap. Then, all the virtual objects, including the occupancy map and VIO, are displayed relative to the $O_w$.

\subsection{Manipulating spatial target} 
%HoloLens offered AR Toolkit(ARTK) \footnote{https://microsoft.github.io/MixedRealityToolkit-Unity/README.html} for building interacting functions. There are four composite gesture enables in HoloLens1. Here we used "Air Tap" and "Tap and hold". 
% \begin{figure}[h]
%     \centering
%     \begin{subfigure}[b]{0.23\textwidth}
%         \centering
%         \includegraphics[width=\textwidth]{ieeeconf/gesture_a.pdf}
%         \caption{}
%     \end{subfigure}
%     \hfill
%     \begin{subfigure}[b]{0.23\textwidth}
%         \centering
%         \includegraphics[width=\textwidth]{ieeeconf/gesture_b.pdf}
%         \caption{}
%     \end{subfigure}
%     \caption{(a) The target drone is highlighted in a white bounding box once head gaze pointed on. (b) Hand gesture is read to select target 3D position after target quadrotor is selected.}
%     \label{fig:my_label}
% \end{figure}

With the Mixed Reality Toolkit (MRTK) \cite{mrtk} offered by Microsoft, our interface allows operators to set the 3D target positions with head gaze and hand gesture. As shown in Fig. \ref{fig:spatialmap}(b), there is a white virtual drone on the 3D map. Similar to $O_w$ selection, the operator moves cursor onto the white drone by head gaze. Then, the virtual drone can be selected by hand gesture and be dragged to the spatial target position. During the process, a text panel, facing the operator, updates the virtual drone position continuously. It is used to assist the operator to set precise position more easily. The function ensures spatial target can be decided intuitively. Since deciding the 3D position is inconvenient to achieve on a flat-screen computer, we present the manipulation method as one of the main contributions of this work.

\subsection{Handling coordinates in HoloLens}
Different objects exist in the AR world, with each having its own coordinate system. We use "position" to represent an object's global position, while "local-position" represents its position in its parent coordinate. The virtual drone, target drone and scaled map shown in Fig. \ref{mrintro} are all rendered under a unified coordinate with origin at $O_w$. Thus, the relative pose between all the virtual objects will not be twisted, even if the operator changes the rendering origin $O_w$. Similarly, the pose panel, belonging to the virtual drone's coordinate, always follow the virtual drone's movement. In our system, HoloLens and the drone have their own independent coordinate.

\subsection{Rendering 3D mesh map}
Besides the occupancy map, the interface can also render a static 3D mesh map, which does not request continuous update. A 3D mesh map uses sliced meshes to represent the 3D space. Each mesh contains the position, normal and color information. After the drone has explored an unknown space, the mesh map is generated offline from the captured images and rendered on AR interface. 
%There are up to 266,000 point clouds rendered at the same time during the actual flight test.

\subsection{Interaction on Rviz}

\begin{figure}[ht]
    \centering
    % \begin{subfigure}[b]{0.23\textwidth}
    %     \centering
    %     \includegraphics[width=\textwidth]{ieeeconf/flightscene.pdf}
    %     \caption{}
    % \end{subfigure}
    % \hfill
    \includegraphics[width=0.3\textwidth]{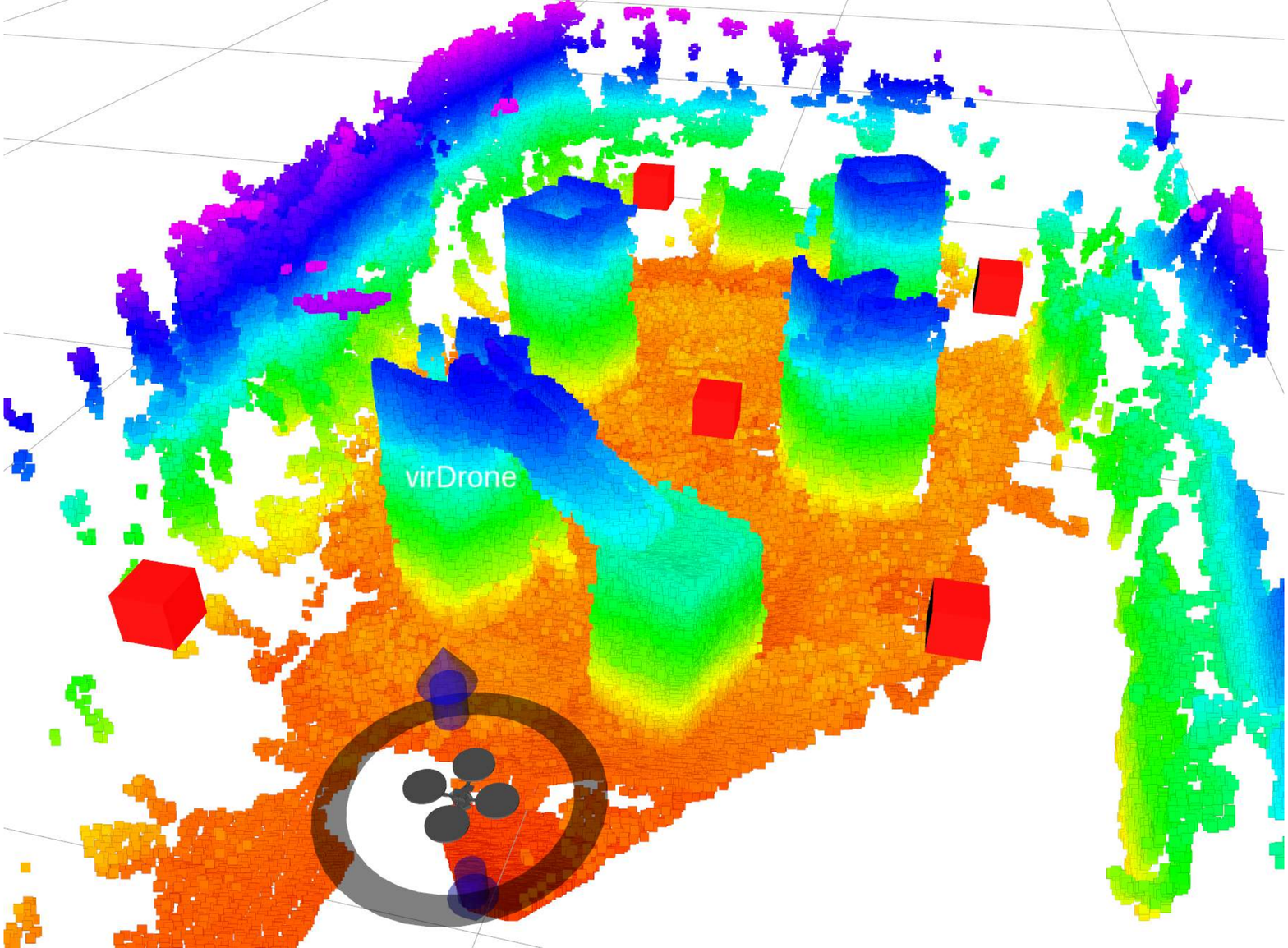}
    \caption{The Rviz interaction interface for comparison.}
    \label{fig:rviz}
\end{figure}

Rviz\footnote{http://wiki.ros.org/rviz} interaction is provided as a comparison. Rviz is a GUI program on ROS, which is running on the ground computer. It can display 3D objects on flat screens, including voxel, pose, and trajectory. The 3D view position and zoom ratio can be changed by a mouse. We develop a Rviz interface, as shown in Fig. \ref{fig:rviz}. It creates a virtual drone with an orbit circle. The operator can click and drag the virtual drone in the space to set a target position. Because of the difficulties of choosing depth on a 2D screen, the orbit circle is used to decide the horizontal position, while the vertical arrow is used for height adjustment. Once the virtual drone is placed correctly, the user clicks a button to send the target command. In this way, the user can distinguish horizontal movement and height movement.

% The Rviz interaction package will be open sourced on Github.

\section{Experiment and Evaluation}

% \subsection{Hardware Preparation}
The onboard computer we choose is a DJI Manifold2-C module. On the ground, there is a desktop computer(Intel i7-8700K CPU and GeForce GTX1080 Ti) running Unity on Windows 10. All of these devices are connected to a wireless router for telecommunication. 
%  The ground computer is connected to the router by CAT6 Ethernet, to eliminate telecommunication time delay.

\subsection{Exploration task}

To validate our interaction approach in an actual navigation mission, we use it to interact with a drone to explore an unknown environment, as shown in Fig.\ref{mrintro}. We set the drone in autonomous navigation mode and continuously set target position via the AR interface. The occupancy map is updated on the AR interface during exploration. Due to the limited computing resources onboard, we save images and odometry on the drone and reconstruct a precise 3D mesh map after the mission has finished. The mesh map is built by Surfel Fusion. During the mission, the drone flies beyond the operator's direct visual range.

\subsection{User study on interaction}
To statistically verify the interaction performance, we design an interaction task and select 10 participants to do a user study. In this user study, each participant needs to do an AR trial (as shown in Fig. \ref{fig:renderpcd}) and Rviz trial (as shown in Fig. \ref{fig:rviz}). Users find a previously built 3D mesh map rendered on a table in the AR trial, and a 3D point cloud map displayed on Rviz. Several cubes are placed to represent pre-designated positions. We set five task positions in each flight trial.  There is no actual flying in this user study so that we can decouple the flight effects and focus on evaluating the interaction method. All ten participants are between 20 and 30 years old and all of them have previous experience in using a 3D GUI on the computer, while four have used AR devices before. The key metrics we use to evaluate each participant's performance are as below:

\begin{enumerate}
\item Composition error is the difference between the pre-designated positions $R_d$ and user-set target positions $R_u$,
	\begin{equation}
	\epsilon_c = \sum_{i\in\omega}d_H(R_d,R_u)
	\end{equation}
where $d_H$ is the Hausdorff distance \cite{huttenlocher1993comparing}, which is used to measure to what extent one set of points lies near another set of points.

% The Hausdorff distance measures the extent to which two sets of points are close. We use this value to measure the accuracy of user-set target points.
\item Completion time $T_c$ is the total time to finish a trial.
\item Command count $N_c$ is the total number of target commands sent by each participant. Since the participant may set the target position several times to reach a desired position precisely, $N_c$ reflects the accuracy and usability of the interaction interface.
%%\item \textbf{NASA TLX} 
\end{enumerate}

The general motivation for the user study is to evaluate user performance in setting target positions.
\subsection{Results}

\subsubsection{Exploration mode result}
The exploration mode is tested successful at our flight scene. Key performances are shown in Table. \ref{table:result1}.
\begin{table}[h]
\label{table_example}
    \begin{center}
    \begin{tabular}{|c|c|c|}
    \hline
    Explored Area & Complete Time & Commands Count \\
    \hline \hline
    5*6*3m & 153 secs & 28 \\
    \hline
    \end{tabular}
    \end{center}
\caption{Key performance in exploration mode}
\label{table:result1}
\end{table}

Special notice is needed on the command count. We set more than necessary number of target positions to explore a small area. That's because we want to capture images from various points and views. Also, revisiting previous positions helps VINS-Fusion optimize its pose graph. So we can reconstruct the 3D environment better on Surfel Fusion. 
\begin{table}[h]
\label{table_example}
    \begin{center}
    \begin{tabular}{|c|c|c|c|}
    \hline
     & Update rate & Data size & Broadcast to AR interface \\
    \hline \hline
    Occupancy map& 10 Hz & 272 MB & Yes\\
    \hline
    FPV images & 30 Hz & 1.39 GB & No\\
    \hline
    \end{tabular}
    \end{center}
\caption{Data size of occupancy map and FPV images}
\label{table:bandwidth}
\end{table}

During the mission, the total data size generated onboard is summarized and shown in table. \ref{table:bandwidth}. The occupancy map's total size is much smaller than the FPV images, and it is not transmitted to the AR interface during the mission. This proves our AR interface requires a smaller communication bandwidth in remote exploration. We also monitor the performance of HoloLens. Its graphics display rate is maintained around 50 Hz and ensures smooth display.

After the drone has landed, a 3D reconstruction is achieved offline and rendered on the AR interface. The results can be found in the experiment video.

\subsubsection{User study result}
We record all users interaction and summarize their key performance indicators by calculating the average value of each metric.
\begin{figure}[ht]
    \centering
    \includegraphics[width= \columnwidth]{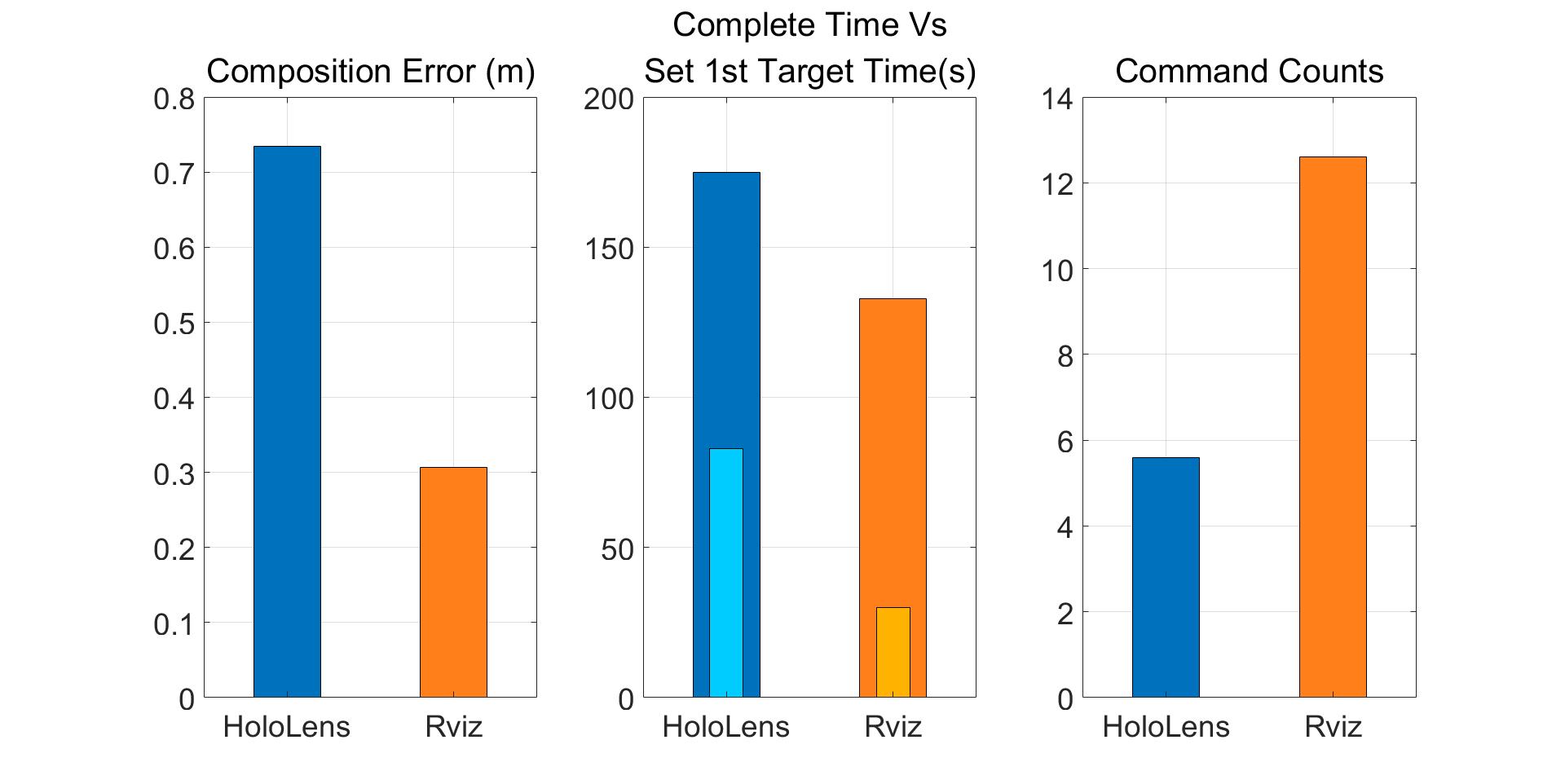}
    \caption{User study result: the highlighted bar in completion time is the time cost for setting the first target.}
    \label{fig:result2}
\end{figure}

As shown in Fig. \ref{fig:result2}, the Rviz interface can set target positions with less error than the AR interface. The result is understandable because current Rviz interface offers axis to set position in each direction, which is an obvious help in setting the precise position. We also find that the AR users normally require a longer time than Rviz users, especially in setting the first target. Some of them suffered in finger tap selection in the beginning. And head gaze was not well utilized, with operators not moving the cursor on the virtual drone and failing in virtual drone selection. We especially highlight the completion time in setting the first target position. The AR interface takes a longer time in setting the first target, while completing later tasks much faster. In terms of command count, although the AR interface allows setting a 3D spatial target in one command, some users still try more than once to reach one target pose.
 
Besides obtaining objective results, we did a brief interview with each user. None of the participants felt physically or mentally uncomfortable in using HoloLens, even though some of them continuously wore the device for 6 minutes. Although the number of participants is not high, we spend a large effort to finish this user study. 
% There are also several user test done on the actual flight, while we exclude them to maintain study consistency.

\subsection{Evaluation}
% \subsubsection{Render Environments Immersive}
Our AR interface \textbf{renders environments immersively}, showing the drone's reconstructed 3D map and real-time odometry information. In exploration mode, the 3D occupancy map from the drone is rendered on a table or the ground near the human operator, who can clearly understand the 3D structure around the drone, realize the localization and controls the viewing angle easily. The interface renders virtual environments smoothly. Although there is a time delay in updating the drone's information, the interface tolerates a time delay without affecting real-time interaction. The final result proves that our interface successfully renders the environment immersively in a way that the human operator can easily understand. However, we found that instructions and practice are necessary for less experienced users to better use the interface. Some users stood statically during the user study because they were used to operate on the ground computer and did not realize they could move to gain a better viewing angle. HoloLens 1 also has the hardware constraint of a very limited field of view(FOV) at 30$^{\circ}$ *17.5$^{\circ}$ (Horizontal * Vertical). This caused frequent complaints that objects were rendered out of the user's field of view. 

% \subsubsection{Intuitive Manipulation}
The second benefit of our interface, as proven by the results, is \textbf{3D Object Manipulation} in setting spatial tasks, which was evaluated by actual flight in exploration mode. Less experienced users also benefit from the simple finger tap target setting. Thanks to the head gaze cursor, most participants were able to select a virtual drone after practice. The command count in Fig. \ref{fig:result2} shows that most participants could set one target position with one command. And all the participants followed their instinct to finger click the virtual drone. However, some of them suffered from moving the virtual drone to the desired 3D position. It is also reflected in the user study result that the AR interface has larger error and longer time. Some participants could not move the drone precisely, causing errors. Several participants suffered from frequently losing track of the selected virtual drone, prolonging completion time. On the other hand, after several practices, some users realized our highlighting function on the virtual drone. So they could understand weather the virtual objects was selected and could manipulate them in the correct way. They obviously took less time in setting the next four target positions. This proves that the manipulation method requires some practice. And it can take much less time to complete the task after experience. In Rviz trial,  the participants felt very comfortable in performing the mission because they were used to a computer GUI like Rviz. Improvements can also be made by adding axis to the virtual objects in the AR interface. Then the operator can move the virtual drone in one direction precisely. In general, we validated the manipulation method, showing that it can be significantly improved after the participant has practiced with it. 

% \subsubsection{Real Flight Application}
Beyond interaction test in user study, we prove this approach can be \textbf{applied in a real autonomous mission}. We combine the interface with our existing autonomous drone system and explored a flying space successfully by our interaction interface. The occupancy map was updated steadily and plenty of target positions were set in a short time. Compare to traditional interface that relies on FPV images, our interface requires a much smaller datalink bandwidth. However, the experiment also inspires us that future work should add an FPV image for the operator to inspect. The data strategy can be optimized to ensure stronger situation awareness at the minimum required bandwidth.

In summary, it has limitations, we have achieved a promising result in applying our AR interface to human-drone interaction for autonomous navigation tasks.

% \subsection{Rviz Performance}
% In using Rviz, on the other hand, the user felt much more comfortable to use it and doesn't require long practice time. That's because 2-D Graphic User Interface is most frequently used nowadays. Since the operator is able to set target by each axis, Rviz achieves the best performance in setting target position precisely. However, the participants on Rviz have to set 2-D target and adjust height, to set a 3D target. This caused longer complete time and larger command frequency. 

\section{Conclusion}
To sum up, this work introduces an Augmented Reality interface that can be remotely used to interact with an autonomous drone. It allows human operator understand the 3D environment around the drone in an immersive way. Intuitive 3D objects manipulation is provided to set 3D targets. The entire system in evaluated in autonomous navigation tasks. Although it is not always perform better than desktop computer interface, we validate its strength in spatial interaction tasks. 

We use HoloLens as an AR device to receive the occupancy map from a drone, and this map is further rendered on physical objects near the operator.  A virtual drone is provided so the operator can set target points by hand gesture and head gaze. It has strength in letting the human operator understand the 3D spatial structure, especially object depth. The interaction requires minimum cognitive knowledge from the user. We evaluated the overall performance by applying the interface on a real autonomous navigation mission. A user study was undertaken and some drawbacks were found. Compared to desktop interface, the AR interface may sacrifice some manipulation accuracy and cost longer completion time. Besides, AR interface requires practice for new users. For future work, more displaying information from drone can be added to the interface, including drone's FPV image and flight status. Thus, the human operator can better interact with autonomous drones in a remote location.

Despite the limitations, this work takes the first step in combining Augmented Reality devices with autonomous navigation of drones.

\addtolength{\textheight}{-12cm}   % This command serves to balance the column lengths

%%%%%%%%%%%%%%%%%%%%%%%%%%%%%%%%%%%%%%%%%%%%%%%%%%%%%%%%%%%%%%%%%%%%%%%%%%%%%%%%

%%%%%%%%%%%%%%%%%%%%%%%%%%%%%%%%%%%%%%%%%%%%%%%%%%%%%%%%%%%%%%%%%%%%%%%%%%%%%%%%



%%%%%%%%%%%%%%%%%%%%%%%%%%%%%%%%%%%%%%%%%%%%%%%%%%%%%%%%%%%%%%%%%%%%%%%%%%%%%%%%

% References are important to the reader; therefore, each citation must be complete and correct. If at all possible, references should be commonly available publications.

\bibliographystyle{unsrt}
\bibliography{lch}

\end{document}